\documentclass[11pt, a4paper]{article}
\pdfoutput=1
\usepackage{jcappub}

\usepackage{booktabs}
\usepackage{multirow}
\usepackage{amssymb}
\usepackage[export]{adjustbox}
\usepackage{floatrow}
\newfloatcommand{capbtabbox}{table}[][3.5in]
\newfloatcommand{capbfigbox}{figure}[][2.3in]
\usepackage{blindtext}
\usepackage{amsmath}
\usepackage{booktabs}
\usepackage{aas_macros}

\newcommand{\ie}{i.e.~}

\def\lsim{\mathrel{\raise.3ex\hbox{$<$\kern-.75em\lower1ex\hbox{$\sim$}}}}
\def\gsim{\mathrel{\raise.3ex\hbox{$>$\kern-.75em\lower1ex\hbox{$\sim$}}}}

\begin{document}

\hspace*{110mm}{\large \tt FERMILAB-PUB-18-254-A}

\vskip 0.2in

\title{Life Versus Dark Energy: How An Advanced Civilization Could Resist the Accelerating Expansion of the Universe}

\author{Dan Hooper}\note{ORCID: http://orcid.org/0000-0001-8837-4127}
\emailAdd{dhooper@fnal.gov}

\affiliation{Fermi National Accelerator Laboratory, Center for Particle Astrophysics, Batavia, IL 60510}
\affiliation{University of Chicago, Department of Astronomy and Astrophysics, Chicago, IL 60637}
\affiliation{University of Chicago, Kavli Institute for Cosmological Physics, Chicago, IL 60637}

\abstract{The presence of dark energy in our universe is causing space to expand at an accelerating rate. As a result, over the next approximately 100 billion years, all stars residing beyond the Local Group will fall beyond the cosmic horizon and become not only unobservable, but entirely inaccessible, thus limiting how much energy could one day be extracted from them. Here, we consider the likely response of a highly advanced civilization to this situation. In particular, we argue that in order to maximize its access to useable energy, a sufficiently advanced civilization would chose to expand rapidly outward, build Dyson Spheres or similar structures around encountered stars, and use the energy that is harnessed to accelerate those stars away from the approaching horizon and toward the center of the civilization. We find that such efforts will be most effective for stars with masses in the range of $M\sim (0.2-1) M_{\odot}$, and could lead to the harvesting of stars within a region extending out to several tens of Mpc in radius, potentially increasing the total amount of energy that is available to a future civilization by a factor of several thousand. We also discuss the observable signatures of a civilization elsewhere in the universe that is currently in this state of stellar harvesting.}

\maketitle

\section{Introduction}

While it may be difficult to predict the detailed behavior of an advanced civilization, it is clear that the objectives of any such system would generically require, or at least benefit from, large quantities of useable energy. With this in mind, Freeman Dyson speculated in his 1960 paper~\cite{1960Sci...131.1667D} that such civilizations would be likely to build structures around stars that are capable of collecting all or most of the light emitted, using this energy and then reemitting the waste heat in the form of high-entropy, infrared or sub-millimeter radiation~\cite{1966ApJ...144.1216S}. Such ``Dyson Spheres'' are not only a staple of science fiction, but have also been the target of many astrophysical searches and other scientific investigations~\cite{2018IJAsB..17..112O,2018ApJ...859...40I,2018arXiv180408351Z,2018arXiv180404157O,2018ApJ...855..110S,2017SoSyR..51..422K,2017ARep...61..347K,2017AmJPh..85...14O,2016arXiv161003219L,2016AJ....152...76V,2016ApJ...825L...5S,2016ApJ...825..155H,2016IJAsB..15..127O,2016arXiv160407844L,2016ApJ...816...17W,2015ApJ...810...23Z,2015A&A...581L...5G,2015ApJS..217...25G,2009ASPC..420..415C,2009ApJ...698.2075C,1998AcAau..42..607T}.

On timescales of tens of billions of years and longer, the expansion of the universe will ultimately limit the ability of an advanced civilization to accumulate and consume useable energy, a fact that has only been exacerbated by the discovery of dark energy~\cite{Riess:1998cb,Perlmutter:1998np}. As space expands, stars and other objects fall beyond the cosmic horizon, making it impossible for them to ever again be observed or otherwise interacted with.\footnote{Strictly speaking, objects do not cross the horizon, but are increasingly redshifted as they approach this boundary. In any case, such objects become invisible and unreachable as a result of the expansion of space.} As dark energy comes to increasingly dominate the total energy density, our universe will enter a phase of exponential expansion, $a(t) \propto e^{H t}$, where $H=H_0 \, \Omega^{1/2}_{\Lambda, 0}$ is the asymptotic value of the Hubble constant in terms of the current Hubble constant, $H_0=67.8$ km/s/Mpc, and the abundance of dark energy, $\Omega_{\Lambda,0}=0.692$~\cite{Ade:2015xua}. Within approximately 100 billion years, all of the matter that is not gravitationally bound to the galaxies that make up our Local Group will become causally disconnected from the Milky Way, falling beyond the limits of our cosmic horizon~\cite{Krauss:1999hj,Loeb:2001dh,Nagamine:2002wi,Nagamine:2003ih}.

In this paper, we speculate about how an advanced civilization would respond to the challenge of living in a universe that is dominated by dark energy. Here we have in mind a civilization that has reached Type III status on the Kardashev scale, which entails the ability to harness the energy produced by stars throughout an entire galaxy~\cite{1964SvA.....8..217K}. Given the inevitability of the encroaching horizon, any sufficiently advanced civilization that is determined to maximize its ability to utilize energy will expand throughout the universe, attempting to secure as many stars as possible before they become permanently inaccessible. To this end, they could build Dyson Spheres or other such structures around the stars that are encountered, and use the energy that is collected to propel those stars toward the center of the civilization, where they will become gravitationally bound and thus protected from the future expansion of space. Broadly speaking, the validity of this conclusion relies only on two modest assumptions, namely that 1) a highly advanced civilization will attempt to maximize its access to usable energy, and that 2) our current understanding of dark energy and its impact on the future expansion history of our universe is approximately correct.

In the following, we will calculate which stars could be effectively harvested in this way. We find that very high-mass stars will often evolve beyond the main sequence before reaching their destination of the central civilization, while very low-mass stars will oftentimes generate too little energy (and thus provide too little acceleration) to avoid falling beyond the horizon. For these reasons, stars with masses in the approximate range of $M\sim (0.2-1) M_{\odot}$ will be the most attractive targets of such an effort. A civilization that begins to expand in the current epoch, traveling at a maximum speed of 10\% (1\%) of the speed of light, could harvest stars in this mass range out to a co-moving radius of approximately 50 Mpc (20 Mpc). Unlike more conventional Dyson Spheres, these structures would not necessarily emit in the infrared or sub-millimeter bands, but would instead use the collected energy to propel the captured stars, providing new and potentially distinctive signatures of an advanced civilization in this stage of expansion and stellar collection.

\section{Gathering Stars from Throughout the Local Universe}

The expansion rate of our universe is described by the Friedmann equations. For the case of a spatially flat universe that is dominated by matter and dark energy (with an equation of state of $w=-1$), the first of these equations can be written as follows:
\begin{equation}
\bigg(\frac{\dot{a}}{a}\bigg)^2 = \frac{8\pi G}{3} \bigg[\frac{\Omega_{M,0}}{a^3}+\Omega_{\Lambda,0}\bigg],
\label{friedmann}
\end{equation}
where $a$ is the scale factor, $\dot{a}$ is its time derivative, $G$ is the Newtonian gravitational constant, and the current abundances of matter and dark energy are given by $\Omega_{M,0}=0.308$ and $\Omega_{\Lambda,0}=0.692$, respectively~\cite{Ade:2015xua}. 

Expanding outward at a speed of $v_{\rm exp}$, a civilization could traverse the following co-moving distance as a function of time:
\begin{equation}
d_{\rm CM}(t) = \int^t_0 \frac{v_{\rm exp} \,dt'}{a(t')}.
\end{equation}
When a star is reached, a Dyson Sphere could be constructed and used to accelerate the surrounded star. If a fraction, $\eta$, of this energy is somehow\footnote{We leave it to the advanced civilization to figure out how exactly this would be accomplished.} transferred into the kinetic energy of the star, its speed after a time, $\Delta t$, would be given by:
\begin{eqnarray}
\label{velocity}
v &=& \bigg(\frac{2  L  \Delta t}{M}\bigg)^{1/2} \\
&\approx& 0.012 \,c \, \bigg(\frac{\Delta t}{1 \, {\rm Gyr}}\bigg)^{1/2}\, \bigg(\frac{L}{L_{\odot}}\bigg)^{1/2} \, \bigg(\frac{M_{\odot}}{M}\bigg)^{1/2} \, \bigg(\frac{\eta}{1}\bigg)^{1/2}, \nonumber
\end{eqnarray}
where $c$ is the speed of light and $L_{\odot}$ and $M_{\odot}$ are the luminosity and mass of the Sun, respectively. Although it would be relatively straightforward to accelerate a star by using a series of reflective surfaces that direct the star's light in one direction (\ie a Shkadov thruster~\cite{1987brig.iafcR....S}), this would yield a velocity no greater than $v = L  \Delta t/Mc \approx 7\times 10^{-5} \,c \, (\Delta t/1 \, {\rm Gyr})(L/L_{\odot}) \, (M_{\odot}/M)$, which is much lower than that yielded by the above expression (for $\eta=1$). By using part of the star's mass (or the mass of another star) as a propellant, however, such high velocities could potentially be attained, while respecting both energy and momentum conservation. In our calculations that follow, we will assume that the advanced civilization in question is able to accelerate stars to the velocities described by Eq.~\ref{velocity}.

In this study, we will focus on stars that are evolving on the main sequence, for which the luminosity is related to the star's mass as follows~\cite{2004sipp.book.....H}:
\begin{eqnarray}
\frac{L}{L_{\odot}} &\approx& 0.028 \, \bigg(\frac{M}{0.4 \,M_{\odot}}\bigg)^{2.3}, \, \, \, \,M < 0.43 M_{\odot} \\
\frac{L}{L_{\odot}} &\approx& \bigg(\frac{M}{M_{\odot}}\bigg)^4, \, \, \,\,\,\,\,\,\, 0.43 M_{\odot} < M < 2 M_{\odot} \nonumber \\
\frac{L}{L_{\odot}} &\approx& 16 \, \bigg(\frac{M}{2 \, M_{\odot}}\bigg)^{3.5}, \, \, \, \,\, M > 2 M_{\odot}. \nonumber
\end{eqnarray}
We also assume that each star remains on the main sequence for a length of time that is given by~\cite{2004sipp.book.....H}:
\begin{equation}
\tau_{_{\rm MS}} \approx 10 \, {\rm Gyr} \times (M_{\odot}/M)^{2.5}.
\label{lifetime}
\end{equation}

Combining the above equations, we arrive at the following:
\begin{eqnarray}
v &\approx& 0.0032 \,c \, \bigg(\frac{\Delta t}{1 \, {\rm Gyr}}\bigg)^{1/2} \, \bigg(\frac{M}{0.4 \,M_{\odot}}\bigg)^{0.65} \, \bigg(\frac{\eta}{1}\bigg)^{1/2}, \, \, \, \,M < 0.43 M_{\odot}\\
v &\approx& 0.012 \,c \, \bigg(\frac{\Delta t}{1 \, {\rm Gyr}}\bigg)^{1/2}\,  \bigg(\frac{M}{M_{\odot}}\bigg)^{1.5} \,\bigg(\frac{\eta}{1}\bigg)^{1/2}, \, \, \,\,\,\,\,\,\, 0.43 M_{\odot} < M < 2 M_{\odot} \nonumber \\
v &\approx& 0.034 \,c \, \bigg(\frac{\Delta t}{1 \, {\rm Gyr}}\bigg)^{1/2}  \, \bigg(\frac{M}{2 \, M_{\odot}}\bigg)^{1.25} \, \bigg(\frac{\eta}{1}\bigg)^{1/2}, \, \, \, \,\, M > 2 M_{\odot}.  \nonumber
\end{eqnarray}
%

\begin{figure}
\includegraphics[width=3.15in,angle=0]{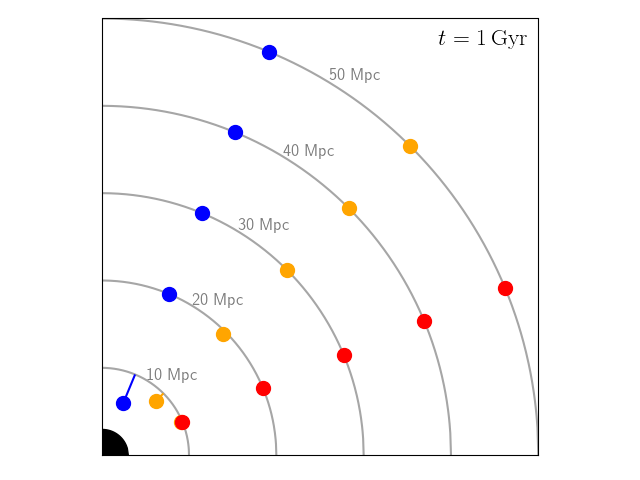}
\hspace{-1.0cm}
\includegraphics[width=3.15in,angle=0]{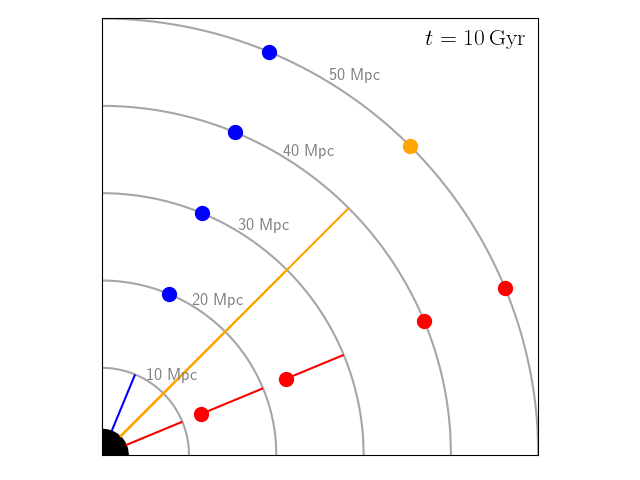}
\hspace{-1.0cm}
\includegraphics[width=3.15in,angle=0]{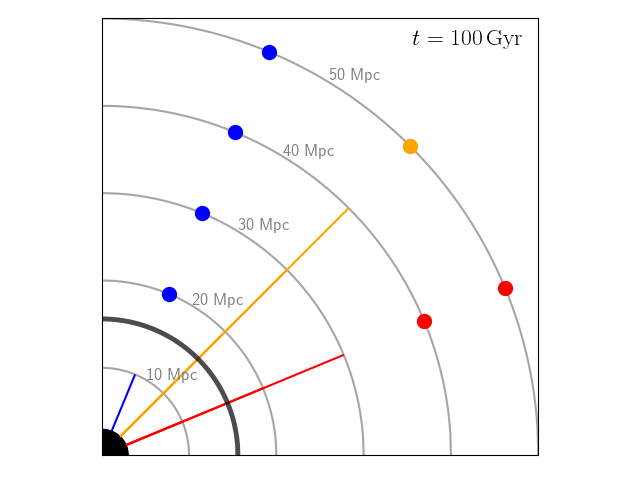}
\caption{A cartoon summarizing the prospects for an advanced civilization to transport usable stars to a central location, assuming that such efforts begin in the present epoch ($t=0$). Blue, yellow and red symbols represent stars with masses of 2, 1 and 0.2$M_{\odot}$, respectively. The colored lines denote the (co-moving) distances those stars have travelled after 1 Gyr, 10 Gyr or 100 Gyr, adopting a maximum speed of 10\% of the speed of light and assuming that approximately 100\% of the collected energy is converted into kinetic energy of the star ($\eta=1$). The results shown apply to the case in which each star is encountered as it begins its main sequence evolution.  The thick black line in the $t=100$ Gyr frame represents the horizon at that time in cosmic history. Very distant stars with either very low or high masses will not be collected, as they will either fall beyond the cosmic horizon or evolve beyond the main sequence before reaching their destination.}
\label{cartoon}
\end{figure}

\begin{figure}
\includegraphics[width=3.08in,angle=0]{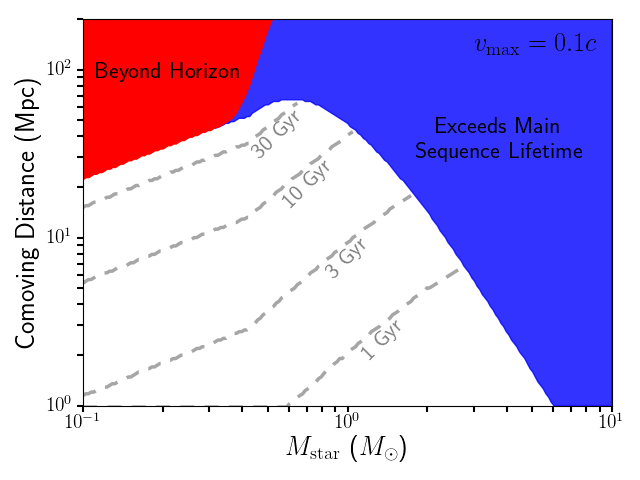}
\hspace{-0.5cm}
\includegraphics[width=3.08in,angle=0]{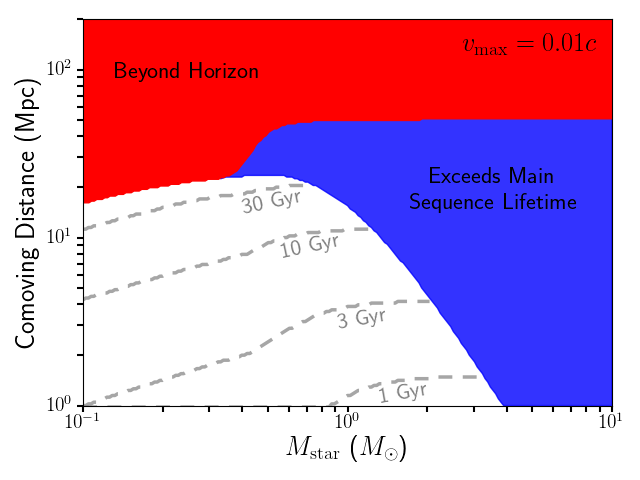}
\hspace{-0.5cm}
\includegraphics[width=3.08in,angle=0]{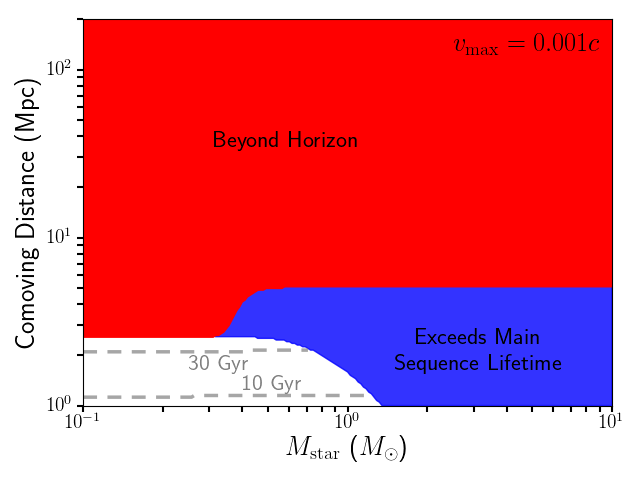}
\caption{A summary of the prospects for an advanced civilization to transport usable stars to a central location, assuming that such efforts begin in the present epoch. Stars in the red (upper left) regions will ultimately fall beyond the cosmic horizon, while those in the blue (right) regions will evolve beyond the main sequence before reaching their destination, and thus not provide useful energy. The grey dashed lines denote the length of time that is required to reach and transport the star. We show results for transport that is limited to speeds below 10\%, 1\% or 0.1\% of the speed of light, and assume that the Dyson Spheres transfer approximately 100\% of the collected energy to the kinetic energy of the star ($\eta=1$). The blue region in each frame has been calculated for the optimistic case of stars that are starting their main sequence evolution at the time that they are encountered (see Fig.~\ref{age}).}
\label{us}
\end{figure}

\begin{figure}
\includegraphics[width=3.08in,angle=0]{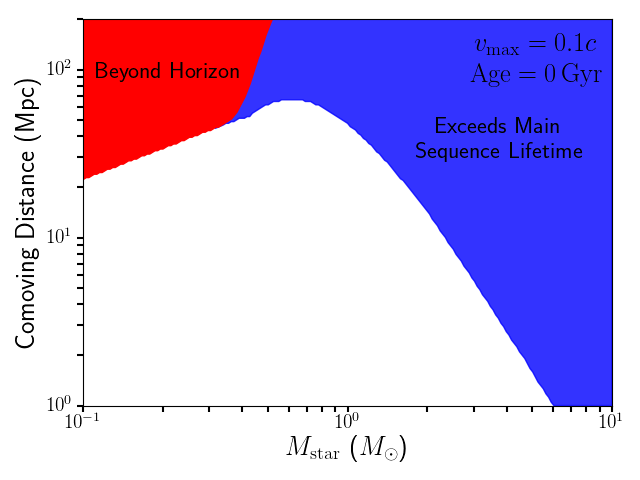}
\hspace{-0.5cm}
\includegraphics[width=3.08in,angle=0]{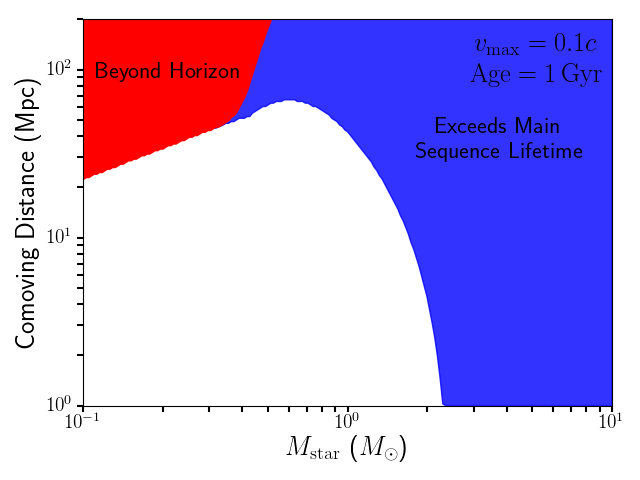}
\hspace{-0.5cm}
\includegraphics[width=3.08in,angle=0]{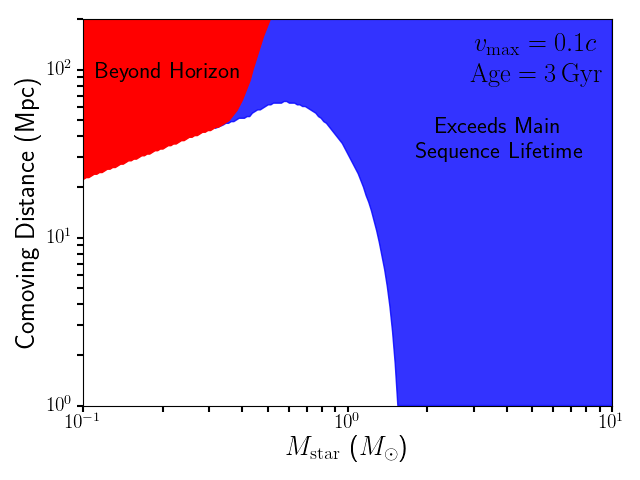}
\hspace{-0.5cm}
\includegraphics[width=3.08in,angle=0]{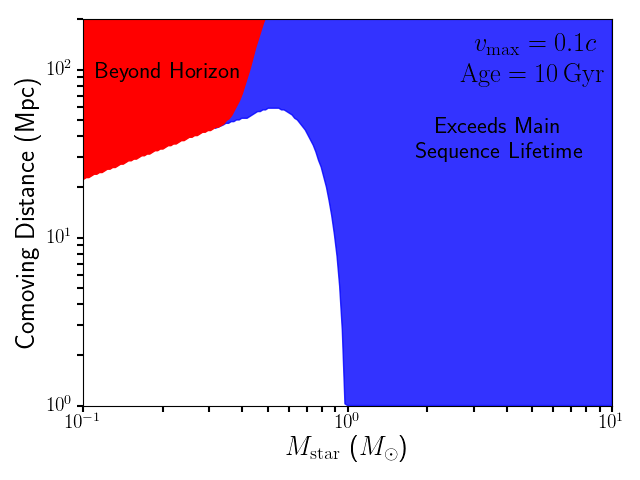}
\caption{As in Fig.~\ref{us}, but for stars that are encountered after having evolved for 1, 3 or 10 billion years on the main sequence. This has little impact on low-mass stars which evolve very slowly, but limits the utility of more massive stars.}
\label{age}
\end{figure}

The ability to transport useable stars to a central civilization will be limited by two factors: 1) the speed of the encroaching horizon, and 2) the lifetime of stars on the main sequence. If an encountered star is either of very low mass (and thus slow to accelerate), very high mass (and thus short lived), and/or very far away, it may be impractical to construct a Dyson Sphere with the goal of transporting it to the central civilization. In Figs.~\ref{cartoon} and~\ref{us}, we summarize the ability of an advanced civilization to transport usable stars to a central location, assuming that such efforts begin in the present epoch. Stars in the red (upper left) region of each frame in Fig.~\ref{us} do not produce enough luminosity to be accelerated to speeds that are fast enough to avoid falling beyond the cosmic horizon. In other words, they will not reach the central civilization within a finite time, and thus will not be the target of such efforts. On the other hand, the stars in the blue (right) regions of this figure will evolve beyond the main sequence prior to arrival at the central location, and thus will not be able to provide useful energy to the civilization in question. More specifically, such stars will reach the central civilization only after a transit time greater than that given in Eq.~\ref{lifetime}. Results are shown in Fig.~\ref{us} for transport that is limited to speeds below 10\%, 1\% or 0.1\% of the speed of light, and in each case we assume that the Dyson Spheres transfer approximately 100\% of the collected energy to the kinetic energy of their captured star ($\eta=1$). In each frame, the grey dashed lines denote the length of time that is required to reach and transport the star. For the case of $v_{\rm max} = 0.1c$, a civilization could ultimately collect and put to use stars that are currently as far away as 65 Mpc.

In calculating the results shown in Figs.~\ref{cartoon} and~\ref{us}, we have optimistically assumed that each star is at the beginning of its main sequence evolution at the time that it is encountered. More realistically, such stars will span a range of ages and stages of their evolution. Such evolution has little impact on low-mass stars which evolve very slowly, but can be quite important for more massive stars. In Fig.~\ref{age}, we show how our results change for the case of stars that are encountered after having evolved for 1, 3 or 10 billion years on the main sequence. 

Integrating our results over the initial mass function of Ref.~\cite{Kroupa:2002ky} and the cosmic star formation rate of Ref.~\cite{Madau:2014bja}, we estimate that an advanced civilization (with $v_{\rm max}=0.1 c$ and $\eta=1$) could increase the total stellar luminosity bound to the Local Group at a point in time 30 billion years in the future by a factor of several thousand relative to that which would have otherwise been available.\footnote{In performing this estimate, we have adopted a density of stars within 1 Mpc of the Local Group that is a factor of 20 greater than the cosmological average~\cite{Gonzalez:2013pqa}.} Over a period of roughly a trillion years, the total luminosity of these stars will drop substantially, but will continue to produce substantial quantities of useable energy due to the longevity of the lightest main sequence stars.

\section{New Observational Signatures of Advanced Civilizations}

\begin{figure}
\includegraphics[width=3.08in,angle=0]{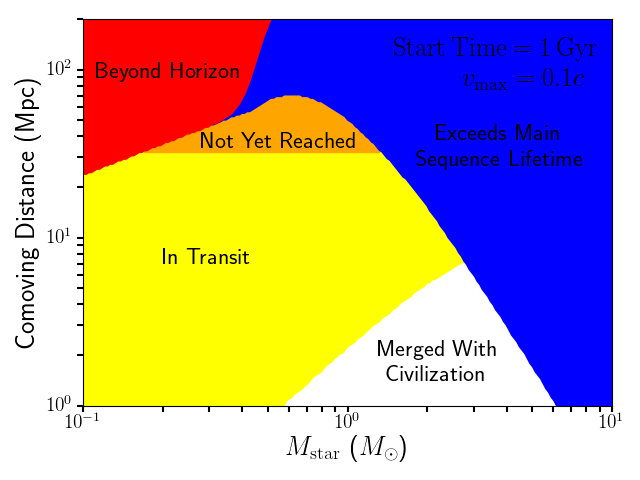}
\hspace{-0.5cm}
\includegraphics[width=3.08in,angle=0]{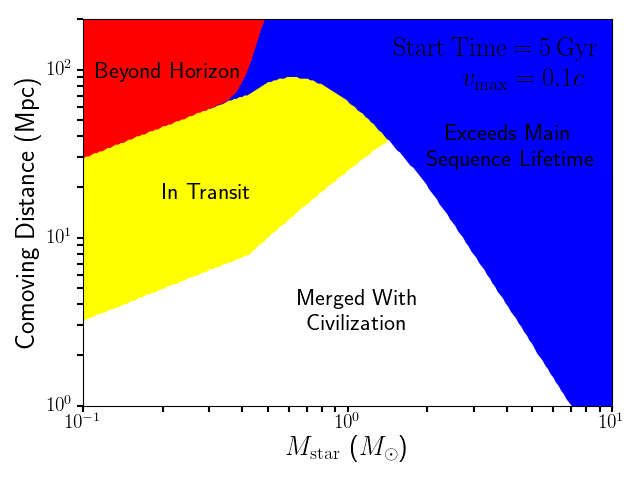}
\caption{As in Fig.~\ref{us}, but for the case of an advanced civilization that began to expand and collect stars 1 Gyr (left) or 5 Gyr (right) in the past. The orange region (only present in the left frame) denotes those stars which have not yet been reached in the present epoch, while those stars in the yellow regions would presently be en route toward the center of the civilization. The stars in the white regions will have already arrived at the central civilization and be providing it with useable energy.}
\label{minus}
\end{figure}

In the previous section, we performed our calculations for the case of an advanced civilization that expands outward from the Milky Way (or Local Group) starting in the current epoch. It is of course possible, however, that life has already evolved elsewhere in our universe, and that civilizations far more advanced than our own may already exist within our Hubble volume. If this is the case, then they may have already begun to collect stars from their surrounding cosmological environment, altering the distribution of stars and leading to potentially observable signatures.

In Fig.~\ref{minus}, we show results for a civilization that began its efforts to collect stars either 1 or 5 billion years ago, in each case adopting a maximum velocity of $v_{\rm max}=0.1 c$ and an efficiency of $\eta=1$.  In the 1 Gyr case, the bulk of the stars within a radius of a few Mpc and with masses in the range of $M \sim (1-4) M_{\odot}$ will have already been collected, while most of the lighter stars out to a radius of $\sim$30 Mpc will presently be in transit on their way to the central civilization. For the 5 Gyr case, stars over a wide range of masses will have already been collected from throughout the surrounding several or even several tens of Mpc. From our vantage point, such a civilization would appear as a extended region, tens of Mpc in radius, with few or no perceivable stars lighter than approximately $\sim$$2 M_{\odot}$ (as such stars will be surrounded by Dyson Spheres). Furthermore, unlike traditional Dyson Spheres, those stars that are currently en route to the central civilization could be visible as a result of the propulsion that they are currently undergoing. The propellant could plausibly take a wide range of forms, and we do not speculate here about its spectral or other signatures. That being said, such acceleration would necessarily require large amounts of energy and likely produce significant fluxes of electromagnetic radiation.

Many of the past searches for Dyson Spheres have focused on detecting the presence of structures around individual stars within the Milky Way (for example, Refs.~\cite{2009ASPC..420..415C,2009ApJ...698.2075C,2018arXiv180408351Z}). Here, we are instead considering galaxies and groups of galaxies in which many or most of the stars are surrounded by Dyson Spheres (and may have been removed from the galaxy), leading to very different observational strategies and signatures~\cite{2015ApJ...810...23Z,2015ApJS..217...25G,2014ApJ...792...27W}. The spectrum of starlight from a galaxy that has had its useful ($M \lsim M_{\odot}$) stars harvested by an advanced civilization would be dominated by massive stars and thus peak at longer wavelengths than otherwise would have been the case. Although such measurements are very challenging, detailed spectroscopy has successfully been used to infer the approximate stellar mass function of nearby galaxies~\cite{2012ApJ...760...71C}.

\section{Summary and Conclusions}

In this paper, we have considered the likely response of a highly advanced civilization to the accelerating expansion of space caused by the presence of dark energy in our universe. If no action is taken, all stars that are not gravitationally bound to the Local Group will move beyond the cosmic horizon and become inaccessible on a timescale of approximately 100 billion years, permanently limiting the quantity of energy that could ultimately be used by such a civilization. In this paper, we argue that in order to maximize its ability to acquire useable energy, a sufficiently advanced civilization will expand rapidly outward, build Dyson Spheres or other similar structures around the stars as they are encountered, and use the collected energy to accelerate the stars away from the encroaching horizon and toward the central civilization. The most attractive targets of such a program will be those stars with masses in the approximate range of $M\sim (0.2-1) M_{\odot}$, as more massive stars are generally too short lived while lighter stars do not produce enough energy to accelerate fast enough to avoid falling beyond the horizon. For a civilization that embarks upon this task in the current epoch, stars in this mass range could be harvested out to a co-moving radius of several tens of Mpc, potentially increasing the total amount of energy that is available in the from of starlight in the distant future by a factor of several thousand.

If an advanced civilization has already embarked upon a program such as this elsewhere in the universe, it would be expected to provide a number of potentially observable signatures. Such a civilization could appear as a region up to tens of Mpc in radius in which most or all of the stars lighter than $\sim 2 M_{\odot}$ are surrounded by Dyson Spheres. Furthermore, during their acceleration and transit toward the central civilization, such stars could be accompanied by visible signals resulting from the large quantities of energy involved in their propulsion.

\bigskip
\bigskip
\bigskip

\textbf{Acknowledgments.} We would like to thank Sam McDermott, Jim Annis and Gordan Krnjaic for very helpful discussions. DH is supported by the US Department of Energy under contract DE-FG02-13ER41958. Fermilab is operated by Fermi Research Alliance, LLC, under contract DE- AC02-07CH11359 with the US Department of Energy.

\bibliography{dyson.bib}
\bibliographystyle{JHEP}

\end{document}